\documentclass[a4paper,11pt]{article}
\usepackage{pos}
\usepackage{axodraw2,pix}
\usepackage{bbm,bm}
\usepackage{slashed}

\renewcommand{\vec}[1]{{\bf #1}}
\newcommand{\rmii}[1]{{\mbox{\tiny\rm{#1}}}}
\newcommand{\gammaE}{{\gamma_\rmii{E}}}

\newcommand{\mD}{m_\rmii{D}}

\newcommand{\lambdaE}{\lambda_\rmii{E}}

\newcommand{\Tint}[1]{{\hbox{$\sum$}\!\!\!\!\!\!\!\int\,}_{\!\!\!\!\raise-0.9ex\hbox{$\scriptstyle{#1}$}}}
\newcommand{\Tinti}[1]{{{\Sigma}\!\!\!\!\raise0.3ex\hbox{$\int$}_\rmii{${#1}$}}}
\newcommand{\Tintip}[1]{{{\Sigma'}\!\!\!\!\!\raise0.3ex\hbox{$\int$}_\rmii{${#1}$}}}

\def\OO{\mathcal{O}}

\def\gsq{g_{\rmii{3d}}^2}
\def\gfour{g_{\rmii{3d}}^4}

\def\mDsq{\mD^2}

\def\mIsqq{m_{\infty,{\rm q}}^2}
\def\mIsqg{m_{\infty,{\rm g}}^2}

\def\Tr{\mathrm{Tr}}

\def\EE{\langle EE \rangle}
\def\BB{\langle BB \rangle}
\def\EB{i \langle EB \rangle}

\def\Zg{Z_{\rm g}}
\def\Zgtd{\Zg^{\rmii{3d}}}
\def\Zf{Z_{\rm f}}

\def\CA{C_\rmii{A}}
\def\CF{C_\rmii{F}}

\def\Nf{N_\rmi{f}}
\def\Tf{T_\rmii{F}}

\def\nB{n_\rmi{B}}
\def\nF{n_\rmi{F}}

\def\TopoWRF(#1,#2){\;\pic{
  #1(-10,15)(30,15)
  #2(15,15)(15,0,180)%
  \GCirc(0,15){2}{0.75}
  \GCirc(30,15){2}{0.75}
 }}
\def\ToptWBBF(#1,#2,#3){\;\pic{
  #1(-10,15)(30,15)
  #2(15,15)(15,0,180)%
  #3(15,15)(7.5,0,180)%
  \GCirc(0,15){2}{0.75}
  \GCirc(30,15){2}{0.75}
  \Vertex(7.5,15){1}
  \Vertex(22.5,15){1}
 }}
 \def\ToptWSBBF(#1,#2,#3,#4,#5){\;\pic{
  #1(-10,15)(30,15)
  #2(15,15)(15,0,70)
  #3(15,15)(15,110,180)
  \GCirc(15,15 70 sin 15 mul add){5}{0.75}
  #4(15,15 70 sin 15 mul add)(5,0,180)
  #5(15,15 70 sin 15 mul add)(5,180,360)%
  \GCirc(0,15){2}{0.75}
  \GCirc(30,15){2}{0.75}
  }}
\def\ToptWSBSBF(#1,#2,#3){\;\pic{
  #1(-10,15)(30,15)
  #2(6,15)(6,0,180)%
  #3(24,15)(6,0,180)%
  \GCirc(0,15){2}{0.75}
  \GCirc(30,15){2}{0.75}
  \Vertex(12,15){1}
  \Vertex(18,15){1}
  }}
\def\ToptWSBBSF(#1,#2,#3){\;\pic{
  #1(-10,15)(30,15)
  #2(10,15)(10,0,180)%
  #3(20,15)(10,180,360)%
  \GCirc(0,15){2}{0.75}
  \GCirc(30,15){2}{0.75}
  \Vertex(10,15){1}
  \Vertex(20,15){1}
  }}
\def\ToptWSF(#1,#2,#3){\;\pic{
  #1(30,15)(-10,15)
  #2(15,15)(15,0,180)
  #3(15,15)(15,180,360)%
  \GCirc(0,15){2}{0.75}
  \GCirc(30,15){2}{0.75}
  }}
\def\ToptWMF(#1,#2,#3,#4){\;\pic{
  #1(-10,15)(30,15)
  #2(15,15)(15,0,90)
  #3(15,15)(15,90,180)
  #4(15,30)(15,15)%
  \GCirc(0,15){2}{0.75}
  \GCirc(30,15){2}{0.75}
  \Vertex(15,15){1}
  }}
\def\ToptWMnF(#1,#2,#3,#4,#5,#6,#7,#8){\;\pic{
  #1(-10,15)(30,15)
  #2(15,15)(15,0,90)
  #3(15,15)(15,90,180)
  #4(15,30)(15,15)%
  \Textt(0,10,#5)
  \Textt(30,10,#6)
  \Textt(15,10,#7)
  \Textb(15,33,#8)
  \GCirc(0,15){2}{0.75}
  \GCirc(30,15){2}{0.75}
  \Vertex(15,15){1}
  }}
\def\ToptWSalF(#1,#2,#3,#4){\;\pic{
  #1(-10,15)(30,15)
  #2(15,15)(15,0,90)
  #3(15,15)(15,90,180)
  #4(0,30)(15,270,360)
  \GCirc(0,15){2}{0.75}
  \GCirc(30,15){2}{0.75}
  }}
\def\ToptWSalnF(#1,#2,#3,#4,#5,#6,#7){\;\pic{
  #1(-10,15)(30,15)
  #2(15,15)(15,0,90)
  #3(15,15)(15,90,180)
  #4(0,30)(15,270,360)
  \Textt(0,10,#5)
  \Textt(30,10,#6)
  \Textb(15,33,#7)
  \GCirc(0,15){2}{0.75}
  \GCirc(30,15){2}{0.75}
  }}
\def\ToptWalF(#1,#2,#3){\;\pic{
  #1(-10,15)(30,15)
  #2(15,15)(15,0,180)
  #3(7.5,15)(7.5,0,180)%
  \GCirc(0,15){2}{0.75}
  \GCirc(30,15){2}{0.75}
  \Vertex(15,15){1}
  }}
\def\ToptWarF(#1,#2,#3){\;\pic{
  #1(-10,15)(30,15)
  #2(15,15)(15,0,180)
  #3(22.5,15)(7.5,0,180)%
  \GCirc(0,15){2}{0.75}
  \GCirc(30,15){2}{0.75}
  \Vertex(15,15){1}
  }}
\def\ToptWTTF(#1,#2,#3){\;\pic{
  #1(-10,15)(30,15)
  #2(0,22.5)(7.5,-90,270)%
  #3(30,22.5)(7.5,-90,270)%
  \GCirc(0,15){2}{0.75}
  \GCirc(30,15){2}{0.75}
 }}
\def\ToptWTRF(#1,#2,#3){\;\pic{
  #1(-10,15)(30,15)
  #2(0,22.5)(7.5,-90,270)%
  #3(22.5,15)(7.5,0,180)%
  \GCirc(0,15){2}{0.75}
  \GCirc(30,15){2}{0.75}
  \Vertex(15,15){1}
 }}
\def\TopoWR(#1,#2){\;\pic{
  #1(0,15)(30,15)
  #2(15,15)(15,0,180)%
  \GCirc(0,15){2}{0.75}
  \GCirc(30,15){2}{0.75}
 }}
\def\ToptWBB(#1,#2,#3){\;\pic{
  #1(0,15)(30,15)
  #2(15,15)(15,0,180)%
  #3(15,15)(7.5,0,180)%
  \GCirc(0,15){2}{0.75}
  \GCirc(30,15){2}{0.75}
  \Vertex(7.5,15){1}
  \Vertex(22.5,15){1}
 }}
\def\ToptWSBB(#1,#2,#3,#4,#5){\;\pic{
  #1(0,15)(30,15)
  #2(15,15)(15,0,70)
  #3(15,15)(15,110,180)
  \GCirc(15,15 70 sin 15 mul add){5}{0.75}
  #4(15,15 70 sin 15 mul add)(5,0,180)
  #5(15,15 70 sin 15 mul add)(5,180,360)%
  \GCirc(0,15){2}{0.75}
  \GCirc(30,15){2}{0.75}
  }}
\def\ToptWSBSB(#1,#2,#3){\;\pic{
  #1(0,15)(30,15)
  #2(6,15)(6,0,180)%
  #3(24,15)(6,0,180)%
  \GCirc(0,15){2}{0.75}
  \GCirc(30,15){2}{0.75}
  \Vertex(12,15){1}
  \Vertex(18,15){1}
  }}
\def\ToptWSBBS(#1,#2,#3){\;\pic{
  #1(0,15)(30,15)
  #2(10,15)(10,0,180)%
  #3(20,15)(10,180,360)%
  \GCirc(0,15){2}{0.75}
  \GCirc(30,15){2}{0.75}
  \Vertex(10,15){1}
  \Vertex(20,15){1}
  }}
\def\ToptWS(#1,#2,#3){\;\pic{
  #1(30,15)(0,15)
  #2(15,15)(15,0,180)
  #3(15,15)(15,180,360)%
  \GCirc(0,15){2}{0.75}
  \GCirc(30,15){2}{0.75}
  }}
\def\ToptWM(#1,#2,#3,#4){\;\pic{
  #1(0,15)(30,15)
  #2(15,15)(15,0,90)
  #3(15,15)(15,90,180)
  #4(15,30)(15,15)%
  \GCirc(0,15){2}{0.75}
  \GCirc(30,15){2}{0.75}
  \Vertex(15,15){1}
  }}
\def\ToptWMn(#1,#2,#3,#4,#5,#6,#7,#8){\;\pic{
  #1(0,15)(30,15)
  #2(15,15)(15,0,90)
  #3(15,15)(15,90,180)
  #4(15,30)(15,15)%
  \Textt(0,10,#5)
  \Textt(30,10,#6)
  \Textt(15,10,#7)
  \Textb(15,33,#8)
  \GCirc(0,15){2}{0.75}
  \GCirc(30,15){2}{0.75}
  \Vertex(15,15){1}
  }}
\def\ToptWSal(#1,#2,#3,#4){\;\pic{
  #1(0,15)(30,15)
  #2(15,15)(15,0,90)
  #3(15,15)(15,90,180)
  #4(0,30)(15,270,360)
  \GCirc(0,15){2}{0.75}
  \GCirc(30,15){2}{0.75}
  }}
\def\ToptWSaln(#1,#2,#3,#4,#5,#6,#7){\;\pic{
  #1(0,15)(30,15)
  #2(15,15)(15,0,90)
  #3(15,15)(15,90,180)
  #4(0,30)(15,270,360)
  \Textt(0,10,#5)
  \Textt(30,10,#6)
  \Textb(15,33,#7)
  \GCirc(0,15){2}{0.75}
  \GCirc(30,15){2}{0.75}
  }}
\def\ToptWal(#1,#2,#3){\;\pic{
  #1(0,15)(30,15)
  #2(15,15)(15,0,180)
  #3(7.5,15)(7.5,0,180)%
  \GCirc(0,15){2}{0.75}
  \GCirc(30,15){2}{0.75}
  \Vertex(15,15){1}
  }}
\def\ToptWar(#1,#2,#3){\;\pic{
  #1(0,15)(30,15)
  #2(15,15)(15,0,180)
  #3(22.5,15)(7.5,0,180)%
  \GCirc(0,15){2}{0.75}
  \GCirc(30,15){2}{0.75}
  \Vertex(15,15){1}
  }}
\def\ToptWTT(#1,#2,#3){\;\pic{
  #1(0,15)(30,15)
  #2(0,22.5)(7.5,-90,270)%
  #3(30,22.5)(7.5,-90,270)%
  \GCirc(0,15){2}{0.75}
  \GCirc(30,15){2}{0.75}
 }}
\def\ToptWTR(#1,#2,#3){\;\pic{
  #1(0,15)(30,15)
  #2(0,22.5)(7.5,-90,270)%
  #3(22.5,15)(7.5,0,180)%
  \GCirc(0,15){2}{0.75}
  \GCirc(30,15){2}{0.75}
  \Vertex(15,15){1}
 }}

\title{Hard parton dispersion in the quark-gluon plasma, non-perturbatively}

\author*[a]{Jacopo Ghiglieri}
\author[b]{Guy D. Moore}
\author[c]{Philipp Schicho}
\author[d]{Niels Schlusser}
\author[a]{Eamonn Weitz}

\affiliation[a]{SUBATECH, Nantes Universit\'e, IMT Atlantique, IN2P3/CNRS,\\
4 rue Alfred Kastler, La Chantrerie BP 20722, 44307 Nantes, France\\}

\affiliation[b]{Institut f\"ur Kernphysik, Technische Universit\"at Darmstadt,\\
Schlossgartenstrasse 2,
D-64289 Darmstadt, Germany}

\affiliation[c]{
Institute for Theoretical Physics, Goethe Universit\"at Frankfurt, 
60438 Frankfurt, Germany
}

\affiliation[d]{Biozentrum, University of Basel,
Spitalstrasse 41,
4056 Basel, Switzerland}

\emailAdd{jacopo.ghiglieri@subatech.in2p3.fr}
\emailAdd{guy.moore@physik.tu-darmstadt.de}
\emailAdd{schicho@itp.uni-frankfurt.de}
\emailAdd{niels.schlusser@unibas.ch}
\emailAdd{eamonn.weitz@subatech.in2p3.fr}

\abstract{The in-medium dispersion of hard partons, encoded in their so-called asymptotic mass, receives large non-perturbative contributions from classical gluons, i.e.\ soft gluons 
with large occupation numbers. 
Here, we discuss how the analytical properties of thermal amplitudes allow for a non-perturbative determination of the infrared classical contribution through
lattice determinations in the dimensionally-reduced 
effective theory of 
hot QCD, EQCD. 
We show how these lattice determinations need to be complemented by perturbative two-loop matching calculations between EQCD and QCD, so that the unphysical (classical) ultraviolet behavior of EQCD is replaced by its proper quantum QCD counterpart. We show how lattice and perturbative EQCD are in good agreement in the UV and present an outlook
on the two-loop quantum QCD contribution.}

\FullConference{HardProbes2023\\
 26-31 March 2023 \\
 Aschaffenburg, Germany\\}

\begin{document}
\maketitle
\section{Introduction}

Jets are a key \textit{hard probe} in the theoretical and 
experimental investigation of the 
Quark-Gluon Plasma (QGP), an exotic state of strongly
interacting matter. Their study provides
important experimental sources of evidence about the nature of
Quantum Chromodynamics (QCD) under extreme conditions --
see \cite{Connors:2017ptx,Cunqueiro:2021wls,Apolinario:2022vzg} 
for recent reviews.

Here, we focus on jets created by light quarks ($u,d,s$) or gluons.
While massless in vacuum, high-energy particles with momentum 
$\boldsymbol{p}$ 
follow the dispersion relation of massive particles when traversing a 
QGP of temperature $T$, $  \omega_{\boldsymbol{p} }^2 = \boldsymbol{p} ^2 + m^2_{\boldsymbol{p}}$,
as found independently by Klimov~\cite{Klimov:1982bv} 
and Weldon~\cite{Weldon:1982aq,Weldon:1982bn}.
This effective mass arises through \textit{forward scattering}
with the medium. 
At very large momentum, it is called the {\em asymptotic mass} and it is 
an important ingredient in 
the determination of medium-induced radiation rates and transport 
coefficients, see~\cite{Ghiglieri:2020dpq} for a review.

To one-loop order, these masses are given by
the gauge  and
the fermion condensates  $\Zg$ and $\Zf$:
\begin{equation}
\label{eff_mass}
\mIsqq = g^2\CF^{ }\Bigl( \Zg^{ } + \Zf^{ } \Bigr)
\,, \qquad
\mIsqg = g^2\CA^{ }\Zg^{ } + 2g^2 \Tf^{ }\Nf^{ }\Zf^{ }
\,,
\end{equation}
where
$\mIsqq$  for quarks and
$\mIsqg$  for gluons.
Here
$\CF=4/3$
and $\CA=3$ 
are the quark and gluon quadratic Casimir,
$\Nf$ is the number of light quark species, and
$\Tf=1/2$ is the trace normalization of the fundamental representation.
The condensates $\Zg$ and $\Zf$
are non-local and
have a gauge-invariant definition
in terms of correlators~\cite{Braaten:1991gm,CaronHuot:2008uw}
\begin{equation}
\label{mom_condensates}
  \Zf \equiv \frac{1}{6}
  \Big\langle
    \overline{\psi} \frac{v_\mu\gamma^\mu}{v\cdot D} \psi
  \Big\rangle
  \;,\qquad
  \Zg \equiv - \frac{1}{8} 
  \Big\langle
    v_{\alpha}^{ } F^{\alpha\mu}
    \frac{1}{(v \cdot D)^2}
    v_{\nu}^{ } F^\nu_{\;\; \mu}
  \Big\rangle
  \;,
\end{equation}
where $v^{\mu}=(1,0,0,1)$ is the  four-velocity of the hard particle
and  $\langle\dots\rangle$ denotes a thermal expectation 
value.
Our conventions are as in \cite{Ghiglieri:2021bom}.

Eq.~\eqref{eff_mass} arises from 
integrating out the energy scale of the jet $p\sim E\gg T$,
truncating at first order in $T/E$ and determining
the matching coefficients at $\OO(g^0)$.
If $E\gtrsim T$,
higher orders in the $T/E$ expansion can become relevant 
beyond one-loop
order and spoil
the factorization into fermionic and bosonic condensates  of~\eqref{eff_mass}.
In coordinate space, the inverse 
of a covariant derivative is an integral over separation of a Wilson
line $U$.  
$\Zg$ is then an integral over
$x^+\equiv(x^0+x^z)/2$
of a correlator of two Lorentz-force insertions
\begin{equation}
     \Zg \equiv  
      -\frac{1}{4} 
      \int_0^\infty \! {\rm d} x^+ x^+
     \,\mathrm{Tr}\,\Bigl\langle
       U(-\infty;x^+)v_\mu^{ } F^{\mu \nu} (x^+)\,U(x^+;0)\,
       v_\rho^{ } F^\rho_{ \, \nu}(0)U(0;-\infty)     \Bigr\rangle\;,
       \label{eq:deffund}
\end{equation}
with $F_{\mu \nu}=F_{\mu\nu}^aT^a$.  $T^a$ is in the 
fundamental representation. So are the Wilson lines,
which source path-ordered gluons along the $x^+$
light-cone direction.
At leading order (LO), one easily finds 
\cite{Klimov:1982bv,Weldon:1982aq,Weldon:1982bn}
\begin{equation}
    \Zg\big\vert^\mathrm{LO}
    =\frac{1}{\pi^2}\int_0^\infty {\rm d}q\,q\,\nB(q)=\frac{T^2}{6},\qquad
     \Zf\big\vert^\mathrm{LO}
    =\frac{1}{\pi^2}\int_0^\infty {\rm d}q\,q\,\nF(q)=\frac{T^2}{12}.
    \label{locond}
\end{equation}
The first correction to this affects $\Zg$ only and does not come
from extra loops but from a more careful evaluation of the infrared (IR).
As $\nB(q\ll T)\approx \frac{T}{q}\gg 1$, soft ($q\sim g T$) gluons
are \textit{classical} and their contribution represents an $\OO(g)$
correction, which can be evaluated by properly accounting for the 
collective effects that arise at that scale. This was done in
\cite{CaronHuot:2008uw}, yielding
the next-to-leading order (NLO) {\em viz.} $\Zg\big\vert^\mathrm{NLO}=
-T\mD/(2\pi)$, where $\mD=g T\sqrt{1+\Nf/6}$ is the Debye mass.
This remarkable result  was obtained, following~\cite{CaronHuot:2008ni}, 
by  exploiting the analytical properties
of thermal amplitudes at space- and light-like separations to 
map the classical contribution to Eq.~\eqref{eq:deffund}
to its counterpart $\Zgtd$ in Electrostatic QCD 
(EQCD). EQCD~\cite{Braaten:1994na,Braaten:1995cm,Braaten:1995jr,Kajantie:1995dw,Kajantie:1997tt}
is a dimensionally-reduced (3D) Effective 
Field Theory (EFT) that describes the Matsubara zero-modes 
of the gauge fields, integrating out all non-zero quark and gluon modes.

This mapping is remarkable: as $\Zgtd$ can be determined 
non-perturbatively in lattice EQCD, this then gives an all-order 
determination of the classical contribution to $\Zg$. In general,
the classical soft and ultrasoft ($q\sim g^2T$) modes are responsible 
for the slow convergence --- or outright breakdown in the latter case 
\cite{Linde:1980ts} --- of thermal perturbation theory. Hence,
the availability of a non-perturbative determination circumvents 
this issue.
In this contribution, we then discuss the necessary steps in lattice 
and continuum EQCD, as well as in perturbative thermal QCD, to 
reach this goal. This is based 
on~\cite{Moore:2020wvy,Ghiglieri:2021bom,wip}, to which we refer for
further details. A similar program has already been 
carried out in~\cite{Moore:2019lgw,Moore:2021jwe}
for the transverse scattering kernel.

\section{Lattice determination}
The continuum EQCD action reads
\begin{equation}
\label{eq:EQCD_cont_action}
S_{\rmii{EQCD},c} = \int_{\vec{x}}\bigg\{
    \frac{1}{2} \Tr\,F_{ij} F_{ij}
  + \Tr\,[D_i, \Phi] [D_i, \Phi]
  + \mDsq \Tr\,\Phi^2
  + \lambdaE^{ }(\Tr\,\Phi^2)^2
\bigg\}
  \,.
\end{equation}
Under dimensional reduction $A^0$ becomes an
adjoint scalar, $A^0\to-i\Phi$,
 $\lambdaE\sim g^4T$ and 
$\gsq\approx g^2T$ is the dimensionful gauge 
coupling~\cite{Kajantie:1997tt}, making
EQCD super-renormalizeable.
The EQCD  contribution to Eq.~\eqref{eq:deffund} is then
\begin{equation}
\label{Zg_EQCD}
\Zgtd =
  \frac{T}{2} \int_0^\infty\!\! {\rm d} L\,L\,\langle FF\rangle,
  \text{ where }
  \langle FF\rangle\equiv \bigl( \EE - \BB - \EB \bigr)
\;,\text{ and } E=D_x\Phi,\;\; B= F_{xz}.
\end{equation}
    $\EE \equiv \frac12
    \bigl\langle (D_x \Phi(L))^a\,\tilde{U}^{ab}_\rmii{A}(L,0)\,(D_x\Phi(0))^b \bigr\rangle$ and similarly for $\BB$ and 
    $\EB$.  
Under dimensional reduction $x^+\to L$, which, together with the rotation of $A^0$ in the complex plane, turns the adjoint Wilson line into a non-unitary 3D equivalent $\tilde{U}^{ab}_\rmii{A}(L,0)$, see~\cite{Moore:2020wvy}.

Lattice EQCD provides us with continuum-extrapolated values for the
three correlators $\EE$, $\BB$ and $\EB$ at discrete 
values of the separation $L$ in the range $0.25<\gsq L<3$ at four
different temperatures ---
see~\cite{Moore:2020wvy,Ghiglieri:2021bom}
for details on discretization, renormalization 
and continuum extrapolation.  
Eq.~\eqref{Zg_EQCD} requires an integral over all
values of $L$.  The IR range at large $\gsq L\gg 1$ can be 
addressed by a fitting ansatz based on the expected exponential
falloff from electrostatic and magnetostatic screening.
At short distances, on the other hand, not only are we limited 
by lattice spacing and increasing discretization effects,
but we need to address the fact that the UV of EQCD,
as for any IR EFT, differs from the UV of QCD, as we now discuss.

\section{Ultraviolet matching}

In the deep UV ($\mD L\ll 1$), we expect perturbative EQCD (pEQCD) to be 
applicable; furthermore,
this is where 
super-renormalizability comes in handy, since it dictates that
\begin{equation}
    \label{EQCD_UV}
    \langle FF\rangle_{\mD L\ll 1}=\frac{c_0}{L^3}+\frac{c_2\gsq}{L^2}
    +\frac{c_4\gfour}{L}+\ldots
    \,,
\end{equation}
with constants $c_i$ arising from  short-distance
pEQCD. Only the first two
terms will be divergent when plugged in Eq.~\eqref{Zg_EQCD}.
$c_0/L^3$ gives rise to a power-law divergence: indeed, the classical
contribution to Eq.~\eqref{locond}, obtained from $\nB(q)\to \frac Tq$,
is linearly UV divergent. These two divergences are opposite and, 
if evaluated in the same scheme, cancel: the UV of EQCD, where
$\mD$ and $\lambdaE$ are negligible, must agree with the IR
of bare 4D QCD. As the bare 4D QCD IR 
is included in Eq.~\eqref{locond},
we can simply subtract $c_0/L^3$ $\forall L$ on the EQCD side. 
A simple calculation finds $c_0=2/\pi$.

At NLO in pEQCD,
$c_2\gsq/L^2$ gives rise to a logarithmic divergence,
namely a $c_2\gsq T \ln (\mu/\mD)/2$ contribution to $\Zgtd$,
with $\mu$ a generic UV regulator. This must  necessarily 
be complemented by a $c_2g^2T^2\ln (T/\mu)/2$ contribution from bare 4D QCD
at $\OO(g^2)$. The latter will be the focus of Sec.~\ref{sec_pert}.

\begin{figure}[t]
\centering
  \includegraphics[width=.5\textwidth]{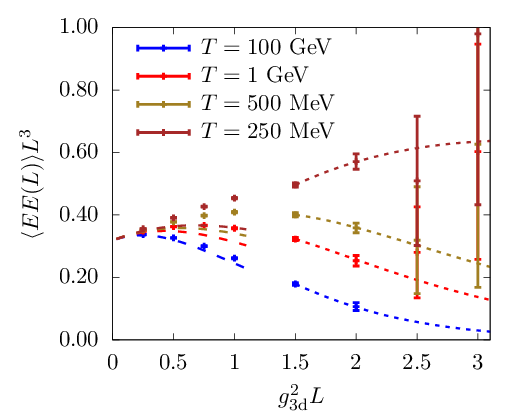}%
  \includegraphics[width=.5\textwidth]{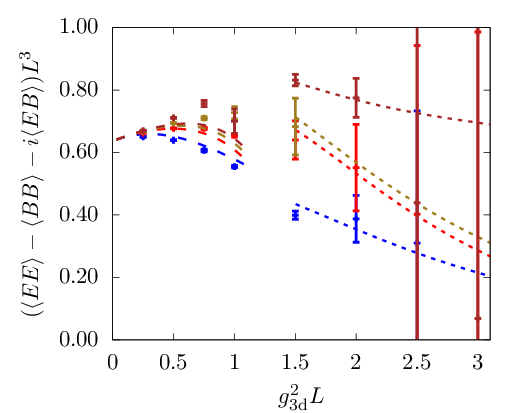}
\caption{
      Continuum-extrapolated
    EQCD lattice data on $\EE$ and
    $\langle FF\rangle$ in eq.~\eqref{Zg_EQCD}
    with
    modelled long $L$-tail (short dashes) and  
    NLO pEQCD (long dashes).  
    $L^3$ compensates the $c_0/L^3$ divergence. Figure 
    from~\cite{Ghiglieri:2021bom}.
    }
\label{fig:comp_LO_NLO_latt}
\end{figure}

In \cite{Ghiglieri:2021bom} we then  
computed $\langle FF\rangle_\mathrm{NLO}$, 
the NLO correction to $\Zgtd$ in pEQCD; expanded 
for $\mD L\ll 1$ this yields $c_2=3/(2\pi^2)$. In 
Fig.~\ref{fig:comp_LO_NLO_latt} we show our lattice data for
the $\EE$ and $\langle FF\rangle$ channels compared 
with the pEQCD evaluation and IR ansatz. As one can see,
pEQCD and lattice agree at the very least at the smallest
separation. We can then merge the two as follows
\begin{align}
    \Zgtd\big\vert^\text{merge} =&\frac{T}{2}\bigg\{ \int_0^{L_\mathrm{min}} {\rm d}L\,L\bigg[
    \langle FF\rangle_\mathrm{NLO}-\frac{c_0}{L^3}-\frac{c_2\gsq}{L^2}\bigg]+\int_{L_\mathrm{min}}^{L_\mathrm{max}} {\rm d}L\,L\bigg[
    \langle FF\rangle_\mathrm{lat}-\frac{c_0}{L^3}\bigg]\nonumber \\
    &
    \hspace{5mm}+\int_{L_\mathrm{max}}^\infty {\rm d}L\,L\bigg[
    \langle FF\rangle_\mathrm{tail}-\frac{c_0}{L^3}\bigg]\bigg\},
    \label{eq_subtr}
\end{align}
where $L_\mathrm{min}$ and $L_\mathrm{max}$ are the smallest and 
largest separations on the lattice and $\langle FF\rangle_\mathrm{lat}$
and $\langle FF\rangle_\mathrm{tail}$ are the lattice and 
IR tail determinations. This procedure can be understood as
splitting the integration 
into three regions: from left to right these are the 
pEQCD, lattice and IR tail ones. As the
$c_2$-term is only subtracted in the perturbative region,
Eq.~\eqref{eq_subtr}
introduces an
artificial logarithmic dependence on its boundary $L_\mathrm{min}$.
Further details and numerical results for this procedure are available 
in \cite{Ghiglieri:2021bom}.

\section{Perturbative determination and outlook}
\label{sec_pert}

\begin{figure}[t]
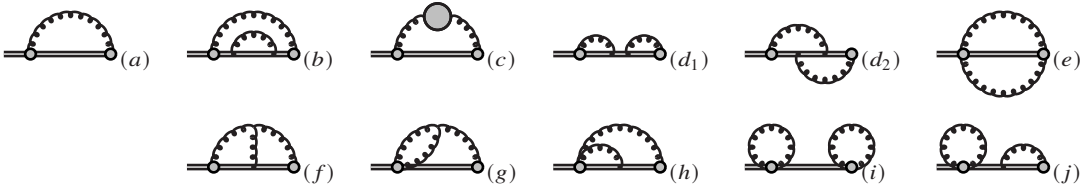

  \centering
  \begin{align*}
    \TopoWRF(\Lsri,\Agliii)_{(a)}\;
    && &
    \ToptWBBF(\Lsri,\Agliii,\Agliii)_{(b)}\;
    &&
    \ToptWSBBF(\Lsri,\Agliii,\Agliii,\Asai,\Asai)_{(c)}\;
    &&
    \ToptWSBSBF(\Lsri,\Agliii,\Agliii)_{(d_1)}\;
    &&
    \ToptWSBBSF(\Lsri,\Agliii,\Agliii)_{(d_2)}\;
    &&
    \ToptWSF(\Lsri,\Agliii,\Agliii)_{(e)}\;
    \nn[3mm]
    && &
    \ToptWMF(\Lsri,\Agliii,\Agliii,\Lgliii)_{(f)}\;
    &&
    \ToptWSalF(\Lsri,\Agliii,\Agliii,\Agliii)_{(g)}\;
    &&
    \ToptWalF(\Lsri,\Agliii,\Agliii)_{(h)}\;
    &&
    \ToptWTTF(\Lsri,\Agliii,\Agliii)_{(i)}\;
    &&
    \ToptWTRF(\Lsri,\Agliii,\Agliii)_{(j)}\;
  \end{align*}
  \vspace{-8mm}
  \caption{%
    Diagrams contributing to $\Zg$ at LO and NLO.
    Gray shaded vertices denote 
    $F^{-\perp}$ insertions;
    a solid line a Wilson line; and
    a curly line a gluon, with the blob its self-energy.
    See~\cite{Ghiglieri:2021bom} for the EQCD counterpart.
  }
  \label{fig:diagrams}
\end{figure}

To lift this artificial cutoff dependence on $L_\mathrm{min}$,
in \cite{wip} we set out to determine the two-loop contribution to $\Zg$ in 
thermal 4D QCD for loop four-momenta of order $T$. In principle this
requires the evaluation of diagrams $(b)$--$(j)$ in Fig.~\ref{fig:diagrams}.
Using the equation of motion of the Wilson line, 
we can show that only diagram $(c)$ contributes in Feynman gauge. 

There we identify the IR-divergent contribution as arising from a thermal
momentum, $K \sim T$ in the blob and a soft one in the outer integration,
$Q\sim g T$. We carefully
evaluate it in dimensional regularization; we do the same for the 
$-c_2\gsq\theta(L_\mathrm{min}-L)/L^2$ subtraction in Eq.~\eqref{eq_subtr}, 
finding that UV and IR poles cancel as expected, yielding
this new matching term~\cite{wip}
\begin{equation}
    \Zg^\mathrm{match} =\frac{ 3g^2  T^2}{4 \pi ^2}\bigg[ \ln (2L_\mathrm{min} T)+ \gammaE -\frac32\bigg]\,,\qquad
    \Zg^\text{non-pert class}=\Zgtd\big\vert^\text{merge}+\Zg^\mathrm{match}\,.
    \label{totmatch}
\end{equation}
The $\ln(L_\mathrm{min})$ dependence vanishes in $\Zg^\text{non-pert class}$,
the sum of Eq.~\eqref{eq_subtr} and $\Zg^\mathrm{match}$. 
$\Zg^\text{non-pert class}$ thus 
provides the complete non-perturbative determination of
the classical contribution to $\Zg$. Numerically, this combination
results in a negative contribution to $\Zg$ which is in magnitude larger
than the negative  $\Zg\big\vert^\mathrm{NLO}=
-T\mD/(2\pi)$. Physically, negativity can be understood as follows.
Let us consider the soft sector of bare 4D QCD. At LO Eq.~\eqref{locond}
implies $\Zg\vert^\text{soft}_\text{bare LO}=\frac{1}{\pi^2}\int_{gT} {\rm d}q\,q\,T/q$. Our procedure amounts to subtracting this bare soft sector
and replacing it 
with its screened EQCD counterpart, which 
reduces its contribution.

Before conclusions on the convergence of this EQCD 
approach can be drawn, it must be noted that $\Zg^\mathrm{match}$
only contains the part of the two-loop 4D QCD calculation responsible
for the IR limit, and that this separation is to some degree arbitrary. 
Only the full determination of the two-loop 4D QCD contribution to $\Zg$
will allow a definite conclusion.\footnote{This is a distinct 
problem from the one recently tackled in 
\cite{Ekstedt:2023anj,Ekstedt:2023oqb,Gorda:2023zwy}, namely the 
two-loop and power corrections to Hard Thermal Loops and their
relation to the asymptotic mass for quasiparticles obeying $T\gg p\gg g T$.}
Preliminary results~\cite{wip}
further indicate the likely emergence of \emph{collinear modes} at this order in perturbation theory. These modes have energies of order $E$
at small angles to $\boldsymbol{p}$, see~\cite{Ghiglieri:2022gyv} for the interplay
of collinear and classical modes in transverse momentum broadening.

\acknowledgments
JG acknowledges support by a PULSAR grant from the R\'egion Pays de la Loire
and by the Agence Nationale de la Recherche under grant ANR-22-CE31-0018 
(AUTOTHERM).
PS was supported
by the European Research Council, grant no.~725369 and
by the Academy of Finland, grant no.~1322507.  PS and GM were supported
by the Deutsche Forschungsgemeinschaft (DFG, German Research Foundation) through the CRC-TR 211 `Strong-interaction matter under extreme conditions' -- project number 315477589 -- TRR 211.

\bibliographystyle{JHEP}
\bibliography{ref.bib}
\end{document}